\documentstyle[prb,aps,epsfig,floats]{revtex}     

\begin{document}

\twocolumn[\hsize\textwidth\columnwidth\hsize\csname@twocolumnfalse\endcsname 
\title{New mechanism of membrane fusion} 
\author{M.\ M\"{u}ller}
\address{Institut f{\"u}r Physik, WA 331, Johannes Gutenberg
Universit{\"a}t,  D-55099 Mainz, Germany}
\author{K.\ Katsov and M.\ Schick}	
\address{Department of Physics, University of Washington, Box 351560, 
Seattle, WA 98195-1560 USA}
\date{\today}
\maketitle
\begin{abstract}
We have carried out Monte Carlo simulation of the fusion of 
bilayers of single chain amphiphiles which show phase behavior 
similar to that of biological lipids. The fusion mechanism we 
observe is very different from the ``stalk'' hypothesis. 
Stalks do form on the first stage of fusion, but they 
do not grow radially to form a hemifused state. Instead,  
stalk formation destabilizes the membranes and results in 
hole formation in the vicinity of the stalks. When holes in 
each bilayer nucleate spontaneously next to the same stalk, 
an incomplete fusion pore is formed. The fusion process 
is completed by propagation of the initial connection, the stalk,
along the edges of the aligned holes. 
\vspace{1.0\baselineskip}
\end{abstract}
]

\section{Introduction}
\vspace{-.8\baselineskip}
Fusion of membranes is involved in basic biological processes but
its mechanism remains poorly understood. That different proteins trigger
different fusion events obscures the possibility of a stage common to
them, one which depends only on the properties of membrane bilayers
themselves.  The existence of such a universality is made plausible by
the realization that, in all fusion processes, membranes must merge, and
the properties of the membranes will most likely determine the nature of
this stage, the fusion intermediate.\cite{ckz}  Time and length
scales of fusion events are of the order of microseconds~\cite{almers}
and nanometers,\cite{ornberg,chandler} so that direct measurements of
intermediate structure have not been possible. Limited theoretical treatment
of this problem, based on  membrane elasticity theory, has focused on the  
so-called {\em stalk} mechanism.\cite{markin,chern,siegel} In this scenario, 
the leaves of the two {\em cis} membranes, closest to one
another, fuse forming a stalk. The stalk expands radially and thins so
that the {\em trans} leaves make contact and form a single bilayer. This
stage is denoted hemifusion. Hole
formation in this bilayer completes the fusion pore.
Due to the lack of direct experimental confirmation, 
this mechanism, although plausible, remains hypothetical.

To obtain a direct view of the fusion mechanism we have undertaken the first
Monte Carlo simulation of membrane fusion, one in which  membranes are
treated on the molecular level. In contrast to the stalk hypothesis, we
find a very different fusion mechanism. Although stalk formation
does take place in the initial stage of fusion, it does not result directly 
in the fusion pore. Instead, the
stalk destabilizes the contacting bilayer membranes and promotes
formation of transient holes {\em next} to it. When  two holes are
nucleated in apposing bilayers in the vicinity of the stalk, 
the latter grows along the hole edges, sealing their periphery like a zipper, 
and eventually closes thereby forming the fusion pore.

\section{The model}
\vspace{-.8\baselineskip}
Our model describes membranes formed by single-chain amphiphiles, like
block copolymers which exhibit the same phases as do biological lipids,
and also form vesicles.\cite{discher} It has the advantage that it has
been well studied, permits detailed analysis of molecular
configurations, and is well suited to processes occurring on the small
time and length scales characteristic of fusion.  The amphiphiles are
treated using the bond fluctuation model~\cite{BFM1} in which each
molecular segment occupies a cube of a three-dimensional lattice.  The
eight lattice sites defining the cube cannot be occupied by another
segment centered on neighboring sites. Segments along an amphiphile are
connected by one of 108 bond vectors of lengths $2,\sqrt{3},\sqrt{5},3$
or $\sqrt{10}$, measured in units of the lattice spacing $a_0$. Mapping
this model onto lipids in solution,\cite{pore} we find the lattice
spacing to correspond to approximately 1\AA. The large number of bond
vectors and the extended segment shape allow a rather faithful
approximation of continuous space, while retaining the computational
advantages of lattice models. The amphiphilic molecules consist of
$N=32$ segments, of which 10 are hydrophilic and 22 are hydrophobic.
This particular choice of the ratio of hydrophilic and hydrophobic segments
results in the diblock system being close to coexistence of the lamellar and
inverted hexagonal phases.
The solvent is represented by  a homopolymer,  
chains consisting of 32 hydrophilic segments.
Like segments attract each other and unlike segments repel
each other via a square well potential which comprises the nearest 54
lattice sites. Each contact changes the energy by an amount
$\epsilon=0.177k_BT$.  The particular choice of the interaction
parameter $\epsilon$ guarantees that the interfacial width between
hydrophilic and hydrophobic segments is not too small to be comparable
to the lattice spacing, and at the same time results in  well-defined
bilayers.

The simulation cell is $L\times L$ in the $x,y$ directions and of length
$D$ in the $z$ direction, with $L=156a_0$ and $D=96a_0$. Periodic
boundary conditions are utilized in all three directions. The monomer
density of the system is $\rho=1/(16a_0^3)$, corresponding to 146,016
segments within the volume, or $2376$ amphiphiles and $2187$ homopolymers.  
To encourage the
fusion of the bilayers that we study, we prepare them under tension by
providing only enough amphiphiles to form bilayers of thickness $25a_0$,
thinner than their equilibrium thickness of $25.2a_0$, which we
determined independently. Given that the system is nearly
incompressible, this corresponds to a fractional increase of area of
less than 0.8\%, one easily sustained by biological
membranes.\cite{evans,needham} Two such bilayers are created parallel to
the $x-y$ plane, stacked one upon the other, thereby mimicking the
dehydration which permits close bilayer contact, a circumstance known to
promote fusion.\cite{cevc}  Sixty-four independent starting
configurations were prepared.  Monte Carlo simulations were performed in
the canonical ensemble.  
The conformations are updated by local monomer displacements and
slithering--snake like movements. The different moves are applied
with a ratio $1:3$. The latter moves do not mimic the realistic dynamics of
lipid molecules and we cannot identify the number of Monte Carlo steps
with time.  However, the density of hydrophilic and hydrophobic segments
is conserved and the molecules diffuse.
Consequently, we expect
the time sequence on length scales larger than a single molecule to
resemble qualitatively those of a simulation with more realistic
dynamics.

To make sure that an isolated membrane is stable, we have simulated a single
bilayer under the above conditions. The system was extremely stable
with respect to  hole formation, apparently due to a very high
energy cost to form the hole's edge, and exhibited the usual capillary
wave fluctuations.
In contrast, the system with two apposing
membranes resulted in a range of structural transformations
that eventually led to fusion pore formation.

\section{Results}
\vspace{-.8\baselineskip}
\subsection{Qualitative observations}
\vspace{-.8\baselineskip}
Initially  the apposed bilayers  formed a number of local connections,
the so-called stalks. Formation of these stalks was promoted by the
system being near 
bulk coexistence between lamellar and inverted hexagonal phases.
In the stalk mechanism, these stalks grow radially
and eventually form  transmembrane contacts. In contrast, we
observed anisotropic growth of the stalks, which resulted in the formation
of structures whose length was a few times that of their width. These
may be precursors of line defects observed
during the lamellar to inverted-hexagonal transition.\cite{siegel}

Formation of the elongated stalks led to  local 
{\em destabilization} of the bilayers. Holes formed next to these defects.
We stress  again that single, isolated, membranes were very stable. Thus 
hole formation in the apposed bilayers is apparently  due to 
inter-bilayer interactions expressed in the stalk formation.
It is easily seen that a hole formed next to a stalk has a smaller 
edge energy, in comparison to
an isolated hole due to the reduction in curvature energy.
This reduction is roughly 
proportional to the length of the edge adjacent to the stalk. 

Hole formation was a transient process, {\em i.e.} numerous
holes were observed to open and close reversibly. 
Only when two holes, one in each bilayer,
nucleated next to, and on the same side of, a given stalk did 
the final stage of the fusion process commence.
The stalk then propagated along the  edges
of the two holes bringing them together, like a zipper, 
to complete the fusion pore.

\begin{figure}[tb]
\begin{center}
\epsfig{file=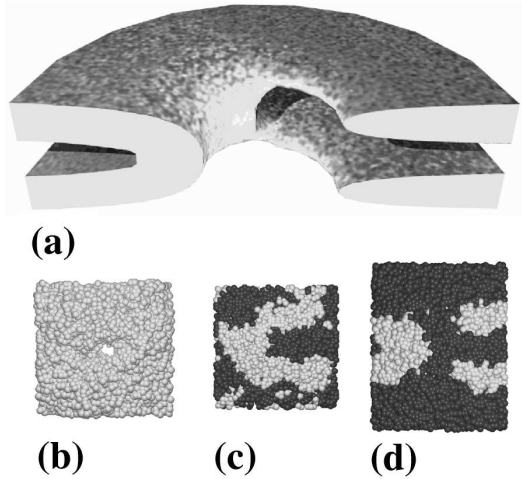,width=2.75in}
\end{center}
\caption{The fusion intermediate. ({\bf a}) Schematic sketch of the 
       intermediate. 
       Only hydrophobic portions are shown.
       ({\bf b}) Top view of a fusion pore through two membranes.  
       Only hydrophobic segments are shown.
       ({\bf c}) Top view of the layer between the two membranes. 
       Hydrophilic segments are dark grey, 
       hydrophobic segments are light grey.
       ({\bf d}) Side view of the snapshot. }
\label{fig_conf}
\end{figure}

We show what one such partially formed pore looks like in
FIG.~\ref{fig_conf}.  
Panel (b) presents the top view of the two membranes. Only tail
segments are shown. The pore through the two bilayers is clearly
visible. Panel (c) presents the top view of the layer between the two
membranes. Hydrophobic segments are light grey, while hydrophilic segments
are dark grey. Connectivity is established along a portion of the pore's rim
by the stalk,
while the two hydrophobic sheets are still unconnected elsewhere.  This
is confirmed by a side view through the pore (d). A schematic of this
intermediate is shown in panel (a). 

\subsection{Quantitative analysis}
\vspace{-.8\baselineskip}
After 62,500 attempted local monomer displacements per segment and
187,500 slithering--snake movements per molecule, we find that the initially sharp
interface of the bilayers have widened, and profiles across the membranes
have adopted locally their stationary form. Profiles at a later stage are
broadened due to fluctuations of the local membrane position,
and to the formation of holes in the membranes. In order to reduce the
effect of membrane fluctuations on our observation of the local
profiles, we define at each  $x$ and $y$ a local midpoint, 
$z_{\rm middle}(x,y)$, of the two-bilayer complex as follows.\cite{block_ana}
For each $x$, $y$ and $z$, we determine the total number of amphiphile segments, $n_a(x,y,z)$, 
in a volume centered at these coordinates, of width $B=28\ a_0$ in the lateral directions, 
and  $50\ a_0$ vertically, the thickness of two bilayers. The value of the
coordinate $z$ which maximizes $n_a(x,y,z)$ 
defines $z_{\rm middle}(x,y)$. We have chosen a rather large
lateral length scale, $B$, so that a small hole in a single bilayer
does not significantly alter our estimate for the midpoint of the
two bilayers. 

To analyze structural changes in the system, we looked not only at 
the fusion pores traversing both bilayers, but also at the holes in each
bilayer which we found to occur about an order of magnitude more frequently
than pores. We did this by calculating the
local density of hydrophobic tail segments coarse grained over $4a_0\times
4a_0$ square columns centered at $x$ and $y$.
Holes and pores should have
a very low, or vanishing, density of tails compared to the areal density
of two normal bilayers.  If the local value of the coarse-grained
density of tail segments within a column extending through the 
whole $z$-range falls
below 25\% of the normal areal density of the two bilayers, 
we define this column as belonging
to a pore traversing them. To locate a hole in only one of the
bilayers, the integration over the $z$ coordinate, rather than
extending over the
whole simulation box width, $D$, is limited to  
either the upper or lower half
of it starting at the point between the two bilayers. If the local
coarse-grained tail density falls below 25\% of the normal areal density
of a single bilayer, the column belongs to a hole.
Pores and holes are defined as the aggregate of nearest neighbor $4a_0
\times 4a_0$ plaquettes with low hydrophobic segment density.

\begin{figure}[tb]
\begin{center}
\epsfig{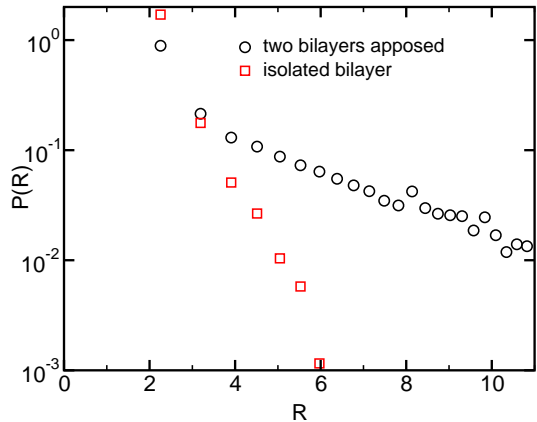}
\end{center}
\caption{Normalized probability distribution of the hole sizes
in a bilayer. Circles refer to the system of two apposed bilayers,
while squares denote the results for a single bilayer. Data for
the two apposed bilayers were collected from 64 systems and 750000 
up to 956250 slithering--snake steps. Data for the single bilayer were sampled over
7 configurations but much longer times.
}
\label{fig_size}
\end{figure}

The probability of hole sizes in the two apposed bilayers and the isolated bilayer 
are presented in FIG.~\ref{fig_size}. Not only is the absolute frequency of hole
formation very much less in the isolated membrane 
than in the two apposed bilayers, but its decay with increasing hole size
is also much more rapid.
This indicates that the edge free energy of holes in the isolated bilayer is larger 
and corroborates the observation that stalk formation promotes hole formation when two
bilayers are in contact.

To determine unambiguously whether fusion pore formation
involves a stage in which the {\em trans} leaves make contact, 
one unavoidable in the standard stalk
mechanism, we perform  a statistical analysis of the fusion intermediates
observed in our simulations.
We examine in each simulation run the first pore which
exceeds a certain radius, $R_{min}=5.5 a_0$, one sufficiently large that
the pore must be open; that is, there is a channel connecting upper and
lower parts of the system accessible to solvent molecules.

Our strategy is to look at the region along the rim, or edge, of
the pore in the pore's mid plane.  If fusion
proceeds {\em via} hemifusion this region has to be essentially
completely hydrophobic.  In contrast, if fusion proceeds through the
formation of two transient holes in apposing bilayers, then we may see pores in an
intermediate stage in which the holes are aligned, but the rim of the
pore has not fully formed. In such a case, the region along the rim of
the pore would be at least partially hydrophilic.  We measure the
density of segments in a series of columns of height $5a_0$ and base
$4a_0\times 4a_0$ centered along the rim of the pore, which is defined
as the points in the $x-y$-mid plane with a distance $14a_0\pm 2a_0$
from the center of the pore. This distance corresponds to a sum of
radius of the pore, $5.5a_0$, and half of the hydrophobic thickness of
the bilayer, $8.5a_0$.

\begin{figure}[tb]
\begin{center}
\epsfig{file=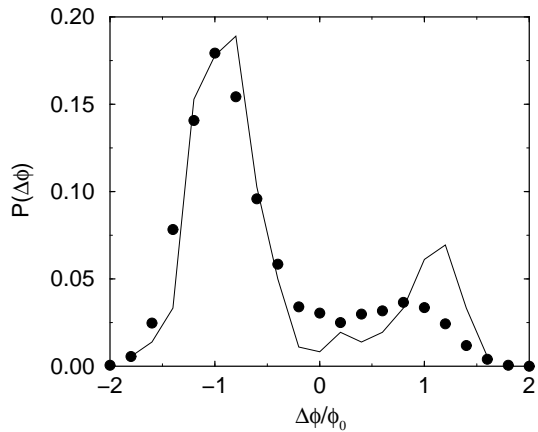,width=2.75in}
\end{center}
\caption{Probability distribution of the difference between hydrophilic and
hydrophobic segment density in the rim.
Positive and negative values correspond to hydrophilic and
hydrophobic dominated regions respectively.
Circles show the average over all 64 independent configurations,
while the solid line depicts the
result for the snapshot in FIG.~\protect\ref{fig_conf}}
\label{fig_distr}
\end{figure}

To quantify the connectivity of the pore's rim, we extract the
distribution of the density difference $\Delta \phi$ between hydrophilic
and hydrophobic segments in this region.  On average there are
$\phi_0=5$ monomers in a column whose volume is $4a_0 \times 4a_0 \times
5a_0$. If the connectivity of the hydrophobic regions of the two
bilayers were complete, as in the stalk mechanism, then we expect the
rim region to be hydrophobic, so that the distribution of $\Delta \phi$
from all these columns would exhibit a single peak around $\Delta
\phi/\phi_0 = -1$. This is not at all what we see. Instead, only 12
pores have this single peaked distribution, whereas the rest are either
bimodal (37 configurations) with a second peak centered about $\Delta
\phi/\phi_0 = +1$ or have a peak at $\Delta \phi/\phi_0 = -1$ with a
long ``tail'' (15 configurations) toward positive $\Delta \phi$. From
this we can conclude that in the majority of cases there is a
considerable portion of the rim which is still hydrophilic; that is,
the rims of the two holes are only partially connected.  The average of
all distributions which we obtain are shown by the solid circles in
FIG.~\ref{fig_distr}. The distribution of the values of
$\Delta \phi$ obtained from the series of columns along the rim of the
particular pore in FIG.~\ref{fig_conf} are shown as a solid line.

\section{Discussion}
\vspace{-.8\baselineskip}
The fusion mechanism we propose is similar to that considered for
electrofusion,\cite{zimmermann} with the addition that we specify the
form of the fusion intermediate. It is in accord with the recent
observation in molecular dynamics simulations that lipid tails take on a
much greater range of configurations, including those sampling the
hydrophilic region, when membranes are brought into close
contact.\cite{satoko}  Our mechanism is similar in spirit to the
initial stages of the mechanism of hemagglutinin-mediated viral fusion
recently proposed by Bonnafous and Stegmann\cite{bonn} who suggest that
before fusion can occur, a hole must first form in the target
membrane. However they go on to suggest that this hole leads to fusion
of the {\em cis} layers producing a hemifusion diaphragm in which a hole
is nucleated. In contrast, hemifusion plays no role in the
scenario we propose. We do observe strong mixing of amphiphiles in the
{\em cis} leaves before pore formation. From such mixing, it is often
inferred that hemifusion precedes complete fusion.\cite{ccgz}  Instead,
we find that this exchange results simply due to formation of the stalks.

The destabilization of the membranes by the stalks that leads to 
formation of the transient holes in both bilayers, observed by us, 
is consistent with the work of Cevc and Richardsen\cite{cevc} who 
emphasize the fact that membrane fusion is strongly promoted by 
{\em defects} in the bilayer structure. 
It seems clear that the additional destabilization of bilayers by the 
defects embedded in the membranes prior to fusion facilitates formation of the
transient holes necessary for fusion.
  
The destabilizing role of the stalk intermediates, observed in our
simulations, is missing from the current phenomenological models
of membrane fusion. Although we do not observe the standard stalk
mechanism, our simulations do not rule it out under
different thermodynamic conditions. Extensions of phenomenological 
approaches that include the new observed intermediates should clarify
this point. We expect the fusion intermediate we have
seen in our simulation to be similar to that in biological membranes. Its
elucidation should facilitate the control and modification of the fusion
process itself.

Note:  While preparing this manuscript, we learned that Noguchi \&
Takasu,\cite{noguchi} studying a very different system consisting of rigid
amphiphilic molecules of three atoms and containing no solvent, observed
fusion behavior over a limited temperature range which is similar to
that we observed in our system. This may indicate a 
universality to the mechanism we have described.  

\begin{acknowledgments}
\vspace{-.8\baselineskip}
Financial support was provided by the National Science
Foundation under grant No. DMR 9876864 and the DFG Bi 314/17 in
the priority program ``wetting and structure formation at interfaces''.
Computer time at the NIC J{\"u}lich, the HLR Stuttgart and the computing
center in Mainz are also gratefully acknowledged.  
\end{acknowledgments}

\vspace{-1.0\baselineskip}

\end{document}